\documentclass[aps,prl,preprint]{revtex4-1}

\usepackage{graphicx}
\usepackage[export]{adjustbox}
\usepackage{amsmath}
\usepackage{float}

\begin{document}

\date{\today}
\newcommand{\Ca}{$^{40}$Ca$^+$}
\newcommand{\CaH}{$^{40}$CaH$^+$}



%
\title{Precision frequency-comb terahertz spectroscopy on pure quantum states of a single molecular ion}

\author{C. W. Chou$^1$, A. L. Collopy$^1$, C. Kurz$^1$, Y. Lin$^{2,1,3,4}$, M. E. Harding$^5$, P. N. Plessow$^6$, T. Fortier$^{1,7}$, S. Diddams$^{1,7}$, D. Leibfried$^{1,7}$, and D. R. Leibrandt$^{1,7}$ }

\affiliation{$^1$Time and Frequency Division, National Institute of Standards and Technology, Boulder, Colorado 80305, USA}

\affiliation{$^2$CAS Key Laboratory of Microscale Magnetic Resonance and Department of Modern Physics, University of Science and Technology of China, Hefei 230026, China}
\affiliation{$^3$Hefei National Laboratory for Physical Sciences at the Microscale, University of Science and Technology of China, Hefei 230026, China
}
\affiliation{$^4$Synergetic Innovation Center of Quantum Information and Quantum Physics, University of Science and Technology of China, Hefei 230026, China
}
\affiliation{$^5$Institute of Nanotechnology, Karlsruhe Institute of Technology, Karlsruhe, Germany}

\affiliation{$^6$Institute of Catalysis Research and Technology, Karlsruhe Institute of Technology, Karlsruhe, Germany}

\affiliation{$^7$University of Colorado, Boulder, Colorado, USA}
\begin{abstract}
Spectroscopy is a powerful tool for studying molecules and is commonly performed on large thermal molecular ensembles that are perturbed by motional shifts and interactions with the environment and one another, resulting in convoluted spectra and limited resolution. Here, we use generally applicable quantum-logic techniques to prepare a trapped molecular ion in a single quantum state, drive terahertz rotational transitions with an optical frequency comb, and read out the final state non-destructively, leaving the molecule ready for further manipulation. We resolve rotational transitions to 11 significant digits and derive the rotational constant of \CaH~to be $B_R =$ 142~501~777.9(1.7)~kHz. Our approach suits a wide range of molecular ions, including polyatomics and species relevant for tests of fundamental physics, chemistry, and astrophysics. 
\end{abstract}
\maketitle
Precision molecular spectroscopy produces information that is essential to understand molecular properties and functions, which underpin chemistry and biology. In particular, microwave rotational spectroscopy can precisely determine various aspects of molecular structure, such as bond lengths and angles, and help identify molecules. However, even in a dilute gas, where the interaction with surrounding molecules is reduced, spectroscopic experiments often fall short of the ultimate resolution set by the natural linewidth of the transitions. This is due to effects that crowd and blur molecular spectra, such as uncontrolled nuclear, rotational, vibrational, and electronic states,  line shifts and broadening from external fields, reduced interaction time from time-of-flight, and the Doppler effect. These limitations have motivated efforts toward trapping molecules and cooling them close to absolute zero temperature. Laser cooling and trapping \cite{neuhauser78,wineland79, phillips98}, which revolutionized atomic physics, have enabled formation of molecules from cold atoms~\cite{ni08} and precision molecular spectroscopy \cite{kondov19}. Direct laser cooling of molecules shows promise for species with advantageous level structures that only require a few laser wavelengths~\cite{shuman10, mccarron18,kozyryev18}, but is infeasible for the vast majority of molecules. Furthermore, even with trapped and cooled molecules~\cite{zeppenfeld12}, commonly used detection methods, such as state-dependent photo-dissociation or ionization~\cite{grossman77, antonov78}, destroy the molecules under study, making them unavailable for further manipulation.

In this work, we perform high resolution spectroscopy on rotational states of a molecular ion using methods that are generally applicable to a broad range of molecular ions, which are readily trapped in electromagnetic potentials~\cite{paul90} and cooled by coupling to co-trapped atomic ions amenable to laser cooling~\cite{molhave00, barrett03}. The long interrogation times and low translational temperature enabled by trapping and sympathetic cooling lead to high resolution~\cite{alighanbari18}, which has, among other advances, enabled the most stringent test of fundamental theory carried out by molecular ions~\cite{cairncross17}. We prepare a trapped \CaH~molecular ion at rest in a single, known quantum state and coherently drive stimulated Raman transitions 
between levels of different rotational quantum numbers $J$, ranging from $J=1$ to $J=6$ in the electronic and vibrational ground state manifold. 
These transitions, with frequencies between 1.4~THz and 3.2~THz, are driven using an optical frequency comb~\cite{hayes10,ding12,leibfried12,solaro18} with a spectrum centered in the range between $800$~nm and $850$~nm, far off-resonance from most vibrational and all electronic transitions~\cite{leibfried12}.  Crucially, the frequency comb spans a large frequency range ($\sim10$~THz) and serves as a versatile and agile tool. It can be used for any allowed two-photon stimulated Raman transition that originates from a substantially thermally occupied state at room temperature in any molecule. We demonstrate $<1$~kHz spectral linewidth and determine the transition centroid frequencies with near 1 part-per-billion (ppb) accuracy. 
The state of the molecule is determined without destroying the molecule~\cite{wolf16,chou17} using quantum-logic spectroscopy (QLS)~\cite{schmidt05}, in which information about the molecular state is transferred to a co-trapped atomic ion. In QLS, the atomic ion acts as a sensitive motion sensor whose state can reflect minute movement of the molecular ion caused by spectroscopic probes. The state of the atomic ion is subsequently detected without perturbing the molecule, which is left in a single quantum state and available for further experiments. Subsequently, it is possible to create superpositions of rotational states or to entangle these states with other quantum mechanical degrees of freedom, such as the internal states of the co-trapped atomic ion. The precision and wealth of data obtainable by this approach could elucidate previously unknown structures and conformations of many molecular ion species. That in turn will provide stringent tests of fundamental physics, and accurate benchmarks useful for improving molecular theories. The spectroscopic information can also help to identify currently unassigned spectral lines observed in the interstellar medium~\cite{mcguire18}. As a first step in this direction, we determine the frequency differences between rotational centroids from measured transition frequencies (see Supplementary Materials) to derive precise \CaH~rotational constants 
up to the fourth order. 

With increasing atom numbers in larger molecules, experimental spectra typically become more complex and the assignment of features to certain transitions is often difficult.  In our experiments, we prepare a nearly pure initial molecular state and detect the final state for the rotational spectra, demonstrating a capability that significantly simplifies the spectra and facilitates assignment of observed lines to specific transitions. In this respect, we also employ quantum-chemical calculations, which have become a useful tool since they often allow prediction of these transitions with enough confidence to allow unambiguous assignment~\cite{RN269}. The calculations yield values for the rotational constants that agree with those derived experimentally.

%
%

\begin{figure}
\includegraphics[angle=0, width=0.95\textwidth]{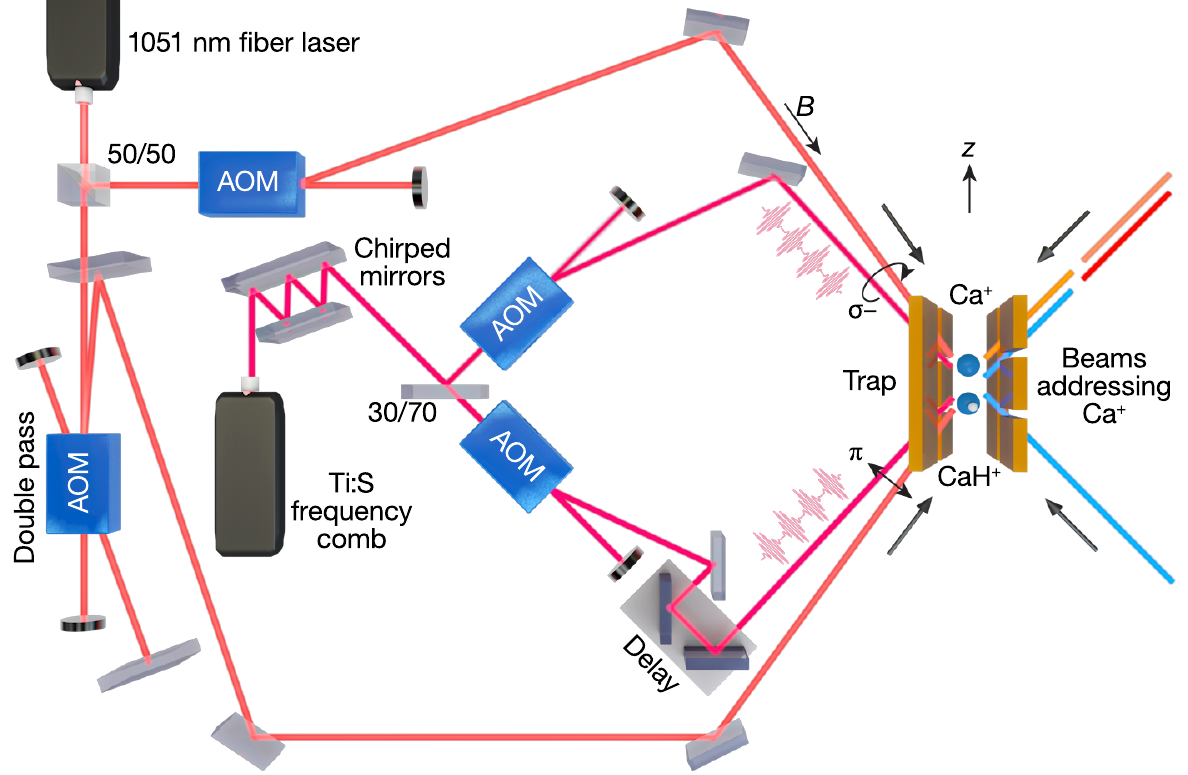}
\caption[Apparatus]{Experimental setup: A \CaH$-$\Ca~ion pair is held in a linear Paul trap. The \CaH~is projectively prepared in pure quantum states using Raman beams from a 1051 nm fiber laser (light red)~\cite{chou17}. The output beam from a Ti:Sapphire (Ti:S) frequency comb is divided into two Raman beams with a 30/70 beamplitter (dark red). The frequencies and power of the beams are controlled by acousto-optic modulators (AOMs). The beams are directed onto the molecular ion at 45 degrees relative to the axis connecting the ions (denoted as $z$). The two pairs of Raman beams have $\pi$ and $\sigma^-$ polarizations respectively relative to the quantization axis defined by the static magnetic field $\mathbf{B}$. They are used to drive two-photon stimulated Raman transitions in the molecule. To ensure that the full spectrum of the comb Raman beams constructively interferes to drive a Raman transition, group delay dispersion (GDD) introduced by optical elements such as AOMs and lenses are pre-compensated with chirped mirrors, and a tunable delay stage ensures that the femtosecond pulses from both arms temporally overlap on the molecular ion.  }
\label{FigSetup}
\end{figure}

The experimental setup is shown in Fig.~\ref{FigSetup}, and a more detailed description of the trap apparatus can be found in~\cite{chou17}. In our experiments, a \Ca$-$\CaH~ion pair is trapped in a linear Paul trap in ultra-high vacuum at room temperature~\cite{chou17}. The \CaH~molecular ion serves as a proof-of-principle test case for a much wider class of molecular ions that could be produced and trapped in a number of ways, but can all be studied with the techniques described here. The \Ca~ion is laser cooled, initialized in a pure quantum state by optical pumping, and coherently manipulated with standard atomic physics methods~\cite{roos99}. It is coupled to the charged molecule by their mutual Coulomb repulsion. The normal modes of translational harmonic motion are shared by both ions and can be sympathetically cooled to their ground state by addressing the \Ca~with suitable laser pulses \cite{barrett03, roos99, rugango15, wolf16}. At room temperature, the \CaH~is in its $\Sigma$ (singlet) electronic and vibrational ground state, but its rotation is in thermal equilibrium with the blackbody radiation from the room temperature environment. Blackbody radiation continuously perturbs the molecule, causing jumps between rotational states on a time scale of tens of milliseconds to seconds for the range of $J$ we consider here.

\begin{figure}
\includegraphics[angle=0, width=1.0\textwidth]{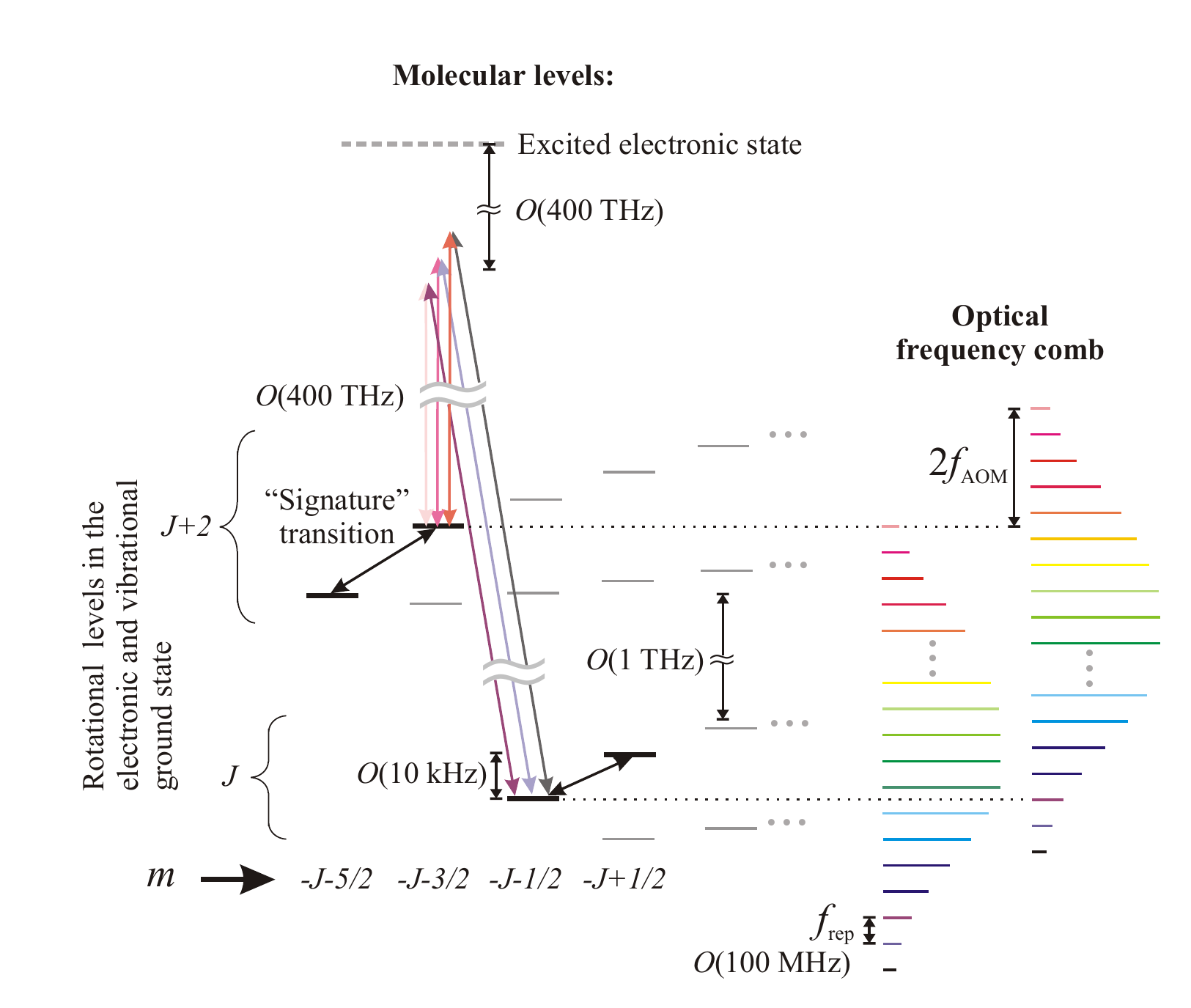}
\caption[Levels]{ Schematic representation of molecular levels probed with comb Raman beams: For \CaH, $\pi$- and $\sigma$-polarized comb Raman beams can induce rotational transitions satisfying $\Delta J=0$, $\pm2$ and $\Delta m=\pm1$.  Within the $J$-th manifold, the $|J,-J-1/2,-\rangle$ and $|J,-J+1/2,-\rangle$ states (black horizontal lines within the same $J$ manifold) connected by the signature transition can be non-destructively detected and prepared with the CW Raman beams. The comb teeth in each beam are equally spaced in frequency by $f_\text{rep}$. Within the limit of the comb spectrum, any comb tooth from one beam can be arranged by the AOMs to have a target difference frequency $f_\text{Raman}$ with a corresponding comb tooth from the other beam. The $\Delta J=\pm 2$ transition $|J,-J-1/2,-\rangle\leftrightarrow|J+2,-J-3/2,-\rangle$ with $J\in\{1,2,3,4\}$ is interrogated by a Raman pulse train from the frequency comb. 
After the interrogation pulse train, the initial and final states can be detected non-destructively. By averaging over many experimental runs, populations can be determined. 
The gray dashed line indicates off-resonant excited electronic states of the molecule. $O$(...) indicates ``on the order of.'' 
}
\label{FigLevelsAndCombDrive}
\end{figure}

Our spectroscopy sequence starts with heralded projective molecular state preparation, as detailed in~\cite{chou17}. We label the eigenstates of the molecule in the presence of a static external magnetic field $\mathbf{B}$ as $|J,m,\xi\rangle$, where $m$ is the projection quantum number of the total (rotational and proton nuclear spin) angular momentum on the direction of $\mathbf{B}$
, and $\xi\in\{+,-\}$ 
labels the two eigenstates that share the same $J$ and $m$, or is the sign of $m$ for the case $m = -J-1/2$ or $J+1/2$~\cite{chou17}. The selection rules $\Delta J = 0,\pm2$ and $\Delta m = \pm1$ apply for stimulated Raman transitions in the linear molecule \CaH~when using a $\sigma^-$-polarized and a $\pi$-polarized light field. For $1\leq J \leq 6$ and the $\sim 0.357$~mT magnetic field used in our experiments, each rotational manifold contains a  ``signature'' transition (Fig.~\ref{FigLevelsAndCombDrive}) with a unique frequency identified from quantum chemistry calculations. Such a signature transition can be specifically targeted with matching probe frequency and used for high-fidelity state preparation and detection for the corresponding manifold~\cite{chou17}. One normal mode of the coupled harmonic motion 
of the atomic and molecular ions in the external potential of the trap is initialized in the ground state $|n=0\rangle$ or the first excited state $|n=1\rangle$ by suitable manipulation of just the atomic ion~\cite{roos99}.  To prepare the molecular state, the molecular population is first pumped towards and concentrated in a state connected by a signature transition~\cite{chou17}. We subsequently attempt to drive the signature transition with a $\pi$ pulse (using a duration on the order of 1~ms such that the transitions are frequency resolved) on a sideband of the motion of the two ions in the trap potential 
$|J,m=-J+1/2,-\rangle|n=0\rangle\leftrightarrow|J,-J-1/2,-\rangle|n=1\rangle$ using a pair of Raman beams derived from a 1051~nm continuous-wave (CW) fiber laser, while tuning their frequency difference with two acousto-optic modulators (AOMs, see Fig.~\ref{FigSetup})~\cite{chou17}. 
With a finite probability, successful projective state preparation is heralded by a change in the motional state, 
which is detected by driving a motional sideband transition between 
the $|S\rangle$ (S$_{1/2}$, magnetic quantum number $m_j = -1/2$, bright in fluorescence detection) and $|D\rangle$ (D$_{5/2}$, $m_j=-5/2$, dark) electronic states of the \Ca~ion that are distinguished by state-dependent fluorescence~\cite{chou17}. With the state of the \Ca~ion prepared in $|D\rangle$, when probing a motion reducing sideband $|D\rangle|n\rangle\leftrightarrow|S\rangle|n-1\rangle$, the atomic state flips when $n>0$, 
while $|D\rangle|n=0\rangle$ remains unchanged, because $n$ cannot be reduced further. Subsequent fluorescence detection on \Ca~constitutes a nearly ideal projective quantum non-demolition molecular state measurement. The projective state preparation can be repeated several times to increase the confidence that the molecule is prepared in the desired state, if successive measurements agree.

 The next steps in our spectroscopic sequence are probing the rotational transition and detecting the molecular state. The CW Raman beams used for projective state preparation and detection can be set to prepare and read out the $|J,-J-1/2,-\rangle$ 
and $|J+2,-J-3/2,-\rangle$ states, each connected by the signature transition in the respective manifold. 
Once the molecule is found in either of these states, we coherently excite the rotational transitions $|J,-J-1/2,-\rangle\leftrightarrow|J+2,-J-3/2,-\rangle$ ($J\in\{1,2,3,4\}$ in this work), with a pair of Raman beams derived from a Titanium:Sapphire (Ti:S) femtosecond laser frequency comb (see Fig.~\ref{FigLevelsAndCombDrive} for the relevant molecular levels). The frequency difference between neighboring comb teeth from these Raman beams can be shifted to any value within the mode spacing (repetition rate) $f_\text{rep}$ of the comb by a pair of AOMs both driven at the frequency $f_\text{AOM}$ 
(see Fig.~\ref{FigSetup} and the Supplementary Materials). The comb teeth in one beam can thus be paired up with the corresponding ones in the other beam and collectively drive a stimulated Raman transition of frequency $f_\text{Raman}=|N f_\text{rep}-2 f_\text{AOM}|$ ($N$ is an integer)  covered by the bandwidth of the frequency comb~\cite{hayes10,ding12,leibfried12,solaro18} (see Fig.~\ref{FigLevelsAndCombDrive} and the Supplementary Materials). In this work, $N$ is in the range between $1.8\times 10^4$ and $4\times10^4$. Subsequently, the molecular state is determined by driving the sidebands of the signature transitions using the CW laser 
followed by motional state detection with the atomic ion. 
By sequentially interrogating the signature transitions in the initial and final $J$ manifolds, we determine whether the attempted transition was successfully made. The experimental sequence of projective state preparation, comb Raman probe, and non-destructive state detection is repeated between 50 and 200 times for each initial state and different comb Raman pulse parameters 
to accumulate statistics on the transition probability 
as a function of the Raman frequency $f_\text{Raman}$ or probe duration.

\begin{figure}
\vspace{-6 cm}
\includegraphics[angle=0, width=1.0\textwidth]{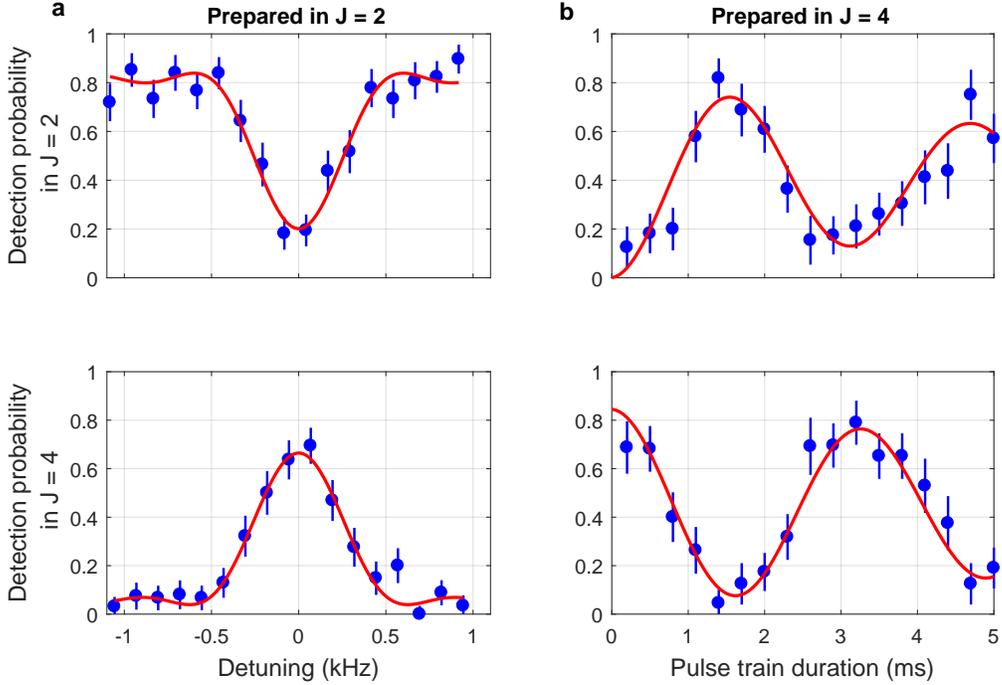}
\vspace{-6 cm}
\caption[Spectra and Rabi flops]{(a) Spectra for the $J = 2 \rightarrow 4$ transition: \CaH~is prepared in  $|J=2,m=-5/2,\xi=-\rangle$, followed by a pulse train from the comb Raman beams probing the $|2,-5/2,-\rangle\leftrightarrow|4,-7/2,-\rangle$ transition. After the probe pulse train, projective measurements of both states are attempted and the state occupation probability is determined from the fraction of successful detections to total number of experiments at this frequency detuning. The probe time is $\sim 1.6$~ms. 
The detection probabilities are mapped out as functions of the Raman detuning. The frequency axis shows 
the offset of the Raman difference frequency from the resonant value. The solid lines are fits to line shapes expected from a square pulse with $\sim 1.6$~ms duration. (b) Rabi flopping on the $J = 2 \leftrightarrow 4$ transition: Starting in  $|J=4,m=-7/2,\xi=-\rangle$, with the comb Raman pulse detuning set to resonance, the state of the \CaH~ion is driven coherently to  $|2,-5/2,-\rangle$ by a pulse train of variable duration from the comb Raman beams. The solid curves are fits to sinusoidal functions with decaying amplitudes. The center wavelength of the frequency comb was $\sim 800$~nm for these spectra and Rabi flopping traces. The error bars stand for $\pm 1$ standard deviation.
}
\label{FigCombSpectraAndFlops}
\end{figure}

Figure~\ref{FigCombSpectraAndFlops} (a) shows the spectra of a transition between the $J = 2$ and $J = 4$ rotational manifolds. When $f_{\text{Raman}}$ of the comb Raman pulse train is tuned near the approximately $2$~THz resonance frequency, the molecular population is transferred from the prepared state $|2,-5/2,-\rangle$ to the final state $|4,-7/2,-\rangle$. The exact value of $N$ of the observed transition is determined by changing $f_\text{rep}$ slightly to $f_\text{rep}+\Delta f_\text{rep}$ and measuring the value of $f_\text{AOM}+\Delta f_\text{AOM}$ that excites the same transition (see Supplementary Materials).  Interrogating molecular ions directly with the frequency comb simplifies the laser setup, because any allowed rotational transition  up to several terahertz can be resonantly interrogated by a frequency comb with 10~THz bandwidth. With the molecule prepared in a known initial state, 
all of the allowed transitions originating from that state can be probed by scanning $f_\text{AOM}$ by $f_\text{rep}/2$, simplifying the search for transitions when knowledge of the molecular constants is limited or unavailable. Performing state detection in both the initial and final states of the transitions confirms the assignment of the observed transitions.

Rabi flopping between the $|J,-J-1/2,-\rangle$ and $|J+2,-J-3/2,-\rangle$ levels is observed by setting $f_\text{Raman}$ on resonance with a transition and varying the duration of the comb Raman pulse train. An example for $J=2$ is shown in Fig.~\ref{FigCombSpectraAndFlops} (b). After the comb Raman pulse train, the molecular ion is in a superposition state of the form $\alpha|J,-J-1/2,-\rangle+\beta|J+2,-J-3/2,-\rangle$, where $\alpha$ and $\beta$ are complex amplitudes ($|\alpha|^2+|\beta|^2=1$). 
The population exchanges approximately sinusoidally between these two levels as a function of pulse duration. Each set of two levels can implement a qubit that can in principle be coherently rotated or entangled with atomic ions or other molecular ions using standard quantum information processing techniques~\cite{blatt08}. The ability to create entangled states of molecules, or atoms and molecules, may enable precision measurements with  quantum advantages on molecular ions. 

With the 20th harmonic of $f_\text{rep}$ phase-locked to a maser-referenced 
1~579~921~905.950~Hz synthesizer, the measured frequencies of transitions with $J$ between 1 and 6 
at $|\mathbf{B}| = 0.357(1)$~mT are presented in Table~\ref{Table1}. For each transition, we characterize the effect of differential AC Stark shifts from the comb Raman beams by measuring the transition frequencies for different intensities while the intensity ratio of $\sigma^-$/$\pi$-polarized light is maintained close to 2~\cite{chou17}, and linearly extrapolating the results to find the zero-intensity transition frequencies. With $<1$~kHz full-width at half-maximum (FWHM) Fourier-limited spectroscopic lineshapes, we reach statistical uncertainties in the line centers below 100~Hz, 
but the trap radio-frequency electric field that drives micro-motion limits the knowledge of line centers at zero electric field to hundreds of hertz in our current setup. A more detailed study of these effects is ongoing and we expect that they can be further suppressed in future experiments. 

\begin{table}
\caption{Measured and inferred rotational transition frequencies. 
The transition frequencies $f_{J_i,J_f}$ were determined at a magnetic field of 0.357(1)~mT with statistical uncertainties $\delta f_{J_i,J_f}$ for the 95~\% confidence intervals of the line centers. The centroid frequencies $cf_{J_i,J_f}$ are calculated from the measured frequencies by subtracting shifts due to finite magnetic field and the spin-rotation coupling. The uncertainties in these corrections and a systematic uncertainty due to the trap radio-frequency electric field at the molecule are reflected in the 95~\%-confidence systematic uncertainties $\delta cf_{J_i,J_f}$ of the centroid that are substantially larger than the statistical uncertainties of the measured resonances. 
}
\begin{tabular}{ccccccc}
\\
\hline
\hline
\multicolumn{1}{c}{$J_i$} & \multicolumn{1}{c}{$J_f$} 
&\multicolumn{1}{c}{ $f_{J_i,J_f}$ (THz)} &\multicolumn{1}{c}{ sta. unc. $\delta f_{J_i,J_f}$ (Hz)} & \multicolumn{1}{c}{ $cf_{J_i,J_f}$ (THz)} & \multicolumn{1}{c}{$\delta cf_{J_i,J_f}$ (kHz)}\\
\hline 
$1$ &$3$         
& 1.424 204 460 565 & 14 & 1.424 204 457 7 & 2.4
\\$2$ &$4$           
& 1.992 911 000 121 & 16 & 1.992 910 990 8 & 3.3
\\$3$&$5$           
& 2.560 643 630 446 & 20 & 2.560 643 614 2 & 3.7
\\$4$&$6$       
&3.127 125 998 610 & 63 & 3.127 125 974 8 & 4.5
\\
\hline
\hline
\end{tabular}
\label{Table1}
\end{table}
We can derive precise values for the rotational constants from the unperturbed rotational transition frequencies. 
 The centroid energy $E_J$ of the $J$th manifold can be parametrized as
\begin{equation}\label{E_J}
E_J = h \sum_{k=1, 2, 3, ...} C_k J^k (J+1)^k \text{, }
\end{equation}
where $h$ is the Planck constant and the expansion coefficients $C_k$ (with units of frequency) correspond to the rotational constant $B_R$ ($k=1$), the centrifugal correction $-D_R$ ($k=2$), the second centrifugal correction $H_R$ ($k=3$), and so on. The inferred frequencies corresponding to the energy differences between the centroids of the rotational manifolds, obtained by subtracting the energy arising from the interactions of the proton and the rotational magnetic moment with the external magnetic field and among themselves~\cite{chou17}, are also listed in Table~\ref{Table1} (see Supplementary Materials). The coefficients $C_k$ in Eq.~\ref{E_J} derived from our measured transition frequencies are shown in Table~\ref{Table2}. The 10-ppb level precision of our \CaH~rotational constant $B_R$ determination is orders of magnitude higher than what can be gained from ab-initio molecular structure calculations. Furthermore, this experiment serves as a proof-of-principle example of a new approach that can be applied to a broad range of molecular ions that were previously inaccessible.

To compute the rotational constants of \CaH, complete basis set extrapolated coupled-cluster calculations at the CCSD(T) level~\cite{ccsdt} were employed in conjunction with incremental corrections for electron correlation up to the CCSDTQ level~\cite{ccsdtq}, as well as relativistic, and diagonal Born-Oppenheimer corrections.
The computed rotational constants in Table~\ref{Table2} are in good agreement with the experimental values. 
The comparison between calculated and experimental results clearly shows that the relative accuracy of computational methods sensitively depends on the computed property. For \CaH, the accuracy of the computed constants is mainly limited by the one-electron basis sets, showing the need for basis set development, basis set extrapolation techniques, and alternative approaches like explicit correlation.

\begin{table}
\caption{Inferred experimental values from four measured rotational transition frequencies and computed values of the molecular constants in Eq.~\ref{E_J}.}
\begin{tabular}{cccc}
\\
\hline
\hline

\multicolumn{1}{c}{$k$} & \multicolumn{1}{c}{Experimental $C_k$ (Hz)} & Computed $C_k$ (Hz)&\multicolumn{1}{c}{Comments}
\\
\hline
 $\text{1  }$        & 1.42 501 777 9 (17) $\times 10^{11}$ &  1.427 (11)  $\times 10^{11}$ &  $B_R$ (rotational constant)
\\ $\text{2  }$        & -5.81217 (19) $\times 10^{6}$      & -5.831 (19)  $\times 10^{6}$  & $-D_R$ (centrifugal correction)
\\ $\text{3  }$        & 222.9 (7.2)                       &  222.6 (0.6)                  & $H_R$ (second centrifugal correction)
\\ $\text{4  }   $        & -0.021 (88)                       & -0.0158(4)                    & Third centrifugal correction
\\
\hline
\hline
\end{tabular}
\label{Table2}
\end{table}

In summary, we drive 1.4~THz to 3.2~THz far off-resonant Raman transitions to interrogate rotational states in a trapped and sympathetically cooled \CaH~molecular ion with ppb-level accuracy and infer state populations and transition rates non-destructively through quantum-logic techniques enabled by a co-trapped atomic ion. The resolution is currently limited by the coherence of the frequency comb that directly drives the transitions and can be further improved~\cite{bartels04}. 
In the future, this approach may enable stringent tests of fundamental physics on a much larger variety of molecular ions than is available by direct laser cooling of charged or neutral molecules, such as searches for variations in the electron-to-proton mass ratio \cite{schiller05,flambaum07} and measurements of minute differences in the transition frequencies of isomers, possibly including chiral molecules with opposite handedness ~\cite{quack00, berger01}. 
When extended to excited vibrational levels, 
the full ro-vibrational energy level structure of a wide range of molecules can be probed, promising rich and accurate information that can benchmark and motivate accurate theoretical models of the potential energy surfaces of molecular ground states. Combined with versatile coherent manipulation enabled by frequency combs, the current protocol could provide new angles for understanding and exploiting molecular dynamics, such as eigenstate analysis of a molecular wavepacket produced by an ultrafast pulse to yield accurate initial states for  studying molecular dynamics. Moreover, coherent manipulation of molecular states may lead to new capabilities, such as alignment and orientation of molecules starting from pure initial states, preparation of squeezed or Schr\"{o}dinger cat-type states of rotation, and precisely state-controlled dissociation of molecular ions.

\section{Acknowledgement}

We thank Flavio C. Cruz and Andrei Kazakov for carefully reading and providing feedback on this manuscript. This work was supported by the US Army Research Office. A. L. C. is supported by a National Research Council postdoctoral fellowship. Y. L. acknowledges support from the National Key R\&D Program of China (Grant No. 2018YFA0306600), the National Natural Science Foundation of China (Grant No. 11974330), Anhui Initiative in Quantum Information Technologies (Grant No. AHY050000) and the Recruitment Program of Global Experts. C. K. acknowledges support from the Alexander von Humboldt foundation. P.N.P. acknowledges support by the state of Baden-W\"{u}rttemberg through bwHPC (bwUnicluster and JUSTUS, RV bw17D01). This is a contribution of the National Institute of Standards and Technology, not subject to US copyright.

\section{Supplementary Materials}
\subsection{The comb Raman beams}
The comb Raman beam 
propagating approximately perpendicular (parallel) to $\mathbf{B}$ is nominally $\pi$-polarized ($\sigma^-$-polarized), with frequency shifted up (down) by the AOM drive frequency $f_{\text{AOM}}$ derived from the same radio-frequency source for both beams. 
The frequency of a comb tooth in the $\pi$-polarized ($\sigma^-$-polarized) comb Raman beam can be expressed as $f_{n_{\pi}}={n_{\pi}} f_{\text{rep}}+f_{\text{CEO}}+f_{\text{AOM}}$ ($f_{n_{\sigma}}=n_{\sigma} f_\text{rep}+f_\text{CEO}-f_\text{AOM}$), where $n_{\pi}$ ($n_{\sigma}$) is a positive integer, $f_{\text{rep}}\simeq 80$~MHz is the repetition rate of the Ti:S laser, and $f_{\text{CEO}}$ is the carrier-envelope-offset frequency. 
The comb Raman beams can resonantly drive Raman transitions with frequencies $f_\text{Raman}=|f_{n_\sigma}-f_{n_\pi}|=|(n_{\sigma}-n_{\pi}) f_\text{rep}-2 f_\text{AOM}|$, 
while angular momentum conservation dictates $\Delta m = \pm 1$, depending on the polarization of the absorbed and stimulated photons. Crucially, there are many comb teeth pairs with equal difference $n_\sigma-n_\pi = N$ that will contribute to driving the transition, as long as the bandwidth of the comb is significantly larger than the corresponding $f_\text{Raman}$ and dispersive phase variations over the comb spectrum are sufficiently small~\cite{solaro18}.

\subsection{Rabi rate of the comb Raman beams and dispersion considerations}
 The Ti:S frequency comb has a center wavelength between 800~nm and 850~nm, and about 20~nm FWHM spectral bandwidth. The average intensities of the two comb beams (labeled $\sigma$ and $\pi$) on the ions are measured via the differential AC Stark shift of the \Ca~$S_{1/2}$ to $D_{5/2}$ 729~nm transition. With the electric field $\mathbf{E}_l$ from the $l$ beam expressed as $\sum_n E_{l,n} \text{Re} (\hat{\mathbf{p}}_l e^{i (\omega_{l,n}t+\phi_{l,n})})$, $l\in\{\sigma,\pi\}$, the on-resonance Rabi rate of the coherent Raman transition $|a\rangle\leftrightarrow|b\rangle$ with frequency $\sim|M f_\text{rep} - 2 f_\text{AOM}|$ driven by the comb beams can be expressed as 
 \begin{eqnarray}
      \Omega_\text{Raman}=\frac{1}{4\hbar^2}\sum_{n,s}\bigg[\frac{\langle b|q\mathbf{r}\cdot \hat{\mathbf{p}}_\pi^* E_{\pi,n}|s\rangle\langle s|q \mathbf{r}\cdot \hat{\mathbf{p}}_\sigma E_{\sigma,n+M}|a\rangle e^{i (\phi_{\pi,n}-\phi_{\sigma,n+M})}}{\omega_{sa}-\omega_{\sigma,n+M}}\nonumber\\
      +\frac{\langle b|q\mathbf{r}\cdot \hat{\mathbf{p}}_\sigma^*E_{\sigma,n+M}|s\rangle\langle s|q \mathbf{r}\cdot \hat{\mathbf{p}}_\pi E_{\pi,n}|a\rangle e^{-i (\phi_{\pi,n}-\phi_{\sigma,n+M})}}{\omega_{sa}+\omega_{\pi,n}}\bigg]\text{.}
 \end{eqnarray}
 The summation runs over all of the Raman paths connecting the initial state $|a\rangle$ via intermediate electronically excited states $\{|s\rangle\}$ to the final state $|b\rangle$. Here $\hat{\mathbf{p}}_\sigma = \frac{1}{\sqrt{2}}(\hat{\mathbf{x}'}-i\hat{\mathbf{y}'})$ and $\hat{\mathbf{p}}_\pi = \hat{\mathbf{z}'}$ are polarization vectors, $\hat{\mathbf{z}'}$ is the unit vector in the direction of $\mathbf{B}$, $\hat{\mathbf{x}'}$ and $\hat{\mathbf{y}'}$ are two unit vectors normal to $\hat{\mathbf{z}'}$ and each other, and $\omega_{l,n} = 2\pi f_{n_l}$, $q$ is the fundamental charge, $\omega_{sa}=(E_s-E_a)/\hbar$ with $E_s$ ($E_a$) the energy of state $|s\rangle$ ($|a\rangle$). 
  The variation in the relative phase between the $n$-th comb tooth in the $\pi$ beam and the ($n+M$)-th comb tooth in the $\sigma$ beam, $\phi_{\pi,n}-\phi_{\sigma, n+M}$, needs to be kept small to ensure constructive interference in the summation. This is achieved by matching the optical path lengths so that the pulses from the two comb beams arrive at the ions at the same time (unmatched propagation distance $\Delta L$ for the beam contributing the $n$-th comb tooth adds an extra $n$-dependent phase difference $2\pi n \frac{f_\text{rep}}{c}\Delta L$), and minimizing the group-delay dispersion (GDD) in the beam paths. The GDD introduced by the optical elements, such as quartz AOMs, 
 focusing lenses, waveplates, and vacuum windows, is pre-compensated with a pair of chirped mirrors at the output of the laser. The single-pass configuration of the AOMs reduces GDD in the comb Raman beams but introduces the complication of variation in beam pointing when varying the AOM drive frequency. This is mitigated by relay-imaging the beam spots in the AOMs onto the ions. To ensure coherence over several milliseconds when driving Raman transitions and the accuracy of frequency measurements, $f_{\text{rep}}$ is tightly phase-locked to a stable direct-digital synthesizer whose frequency is referenced to a hydrogen maser.
 
\subsection{Determination of absolute frequencies of the rotational transitions}
When tuning the frequency of one comb beam relative to the other, the excitation spectrum of the molecule repeats every $f_\text{rep}$. We can still find the absolute rotational transition frequencies by doing two measurements with different $f_\text{rep}$. When $f_\text{rep}$ is changed to $f_\text{rep}+\Delta f_\text{rep}$, $f_\text{AOM}$ needs to be changed to $f_\text{AOM}+\Delta f_\text{AOM}$ to drive the same transition. Thus $N f_\text{rep} -2 f_\text{AOM} = N (f_\text{rep}+\Delta f_\text{rep})-2(f_\text{AOM}+\Delta f_\text{AOM})$ (for the transitions under consideration, $N\equiv n_{\sigma}-n_{\pi}>0$ and $\Delta m = -1$), and the positive integer $N = \frac{2 \Delta f_\text{AOM}}{\Delta f_\text{rep}}$. With narrow linewidth spectra (small uncertainty in $f_\text{AOM}$), the uncertainty in $N$, $\delta N = \sqrt{[\frac{2 \delta(\Delta f_\text{AOM})}{\Delta f_\text{rep}}]^2+[\frac{2\Delta f_\text{AOM}}{\Delta f_\text{rep}^2}\delta(\Delta f_\text{rep})]^2}$ can be reduced to $\ll 1$ with sufficiently large $\Delta f_\text{rep}$ and thus yield the exact integer value of $N$.

\subsection{Determination of the rotational constants}
The energy of the centroid for the $J$-th rotational manifold in terms of the rotational constants $C_k$ is described by Eq.~(\ref{E_J}). The energy difference between the magnetic sublevels $|J,m,\xi\rangle$ and the centroid, is given by the Breit-Rabi formula for the eigenvalues of the spin-rotation Hamiltonian of the molecule in the static external field $\mathbf{B}$~\cite{chou17}: 
\begin{equation}
\hat{H}_{J} = \frac{1}{\hbar} \bigg( -g_J \mu_{N}\hat{\mathbf{J}}\cdot \mathbf{B}-g_I \mu_{N}\hat{\mathbf{I}}\cdot \mathbf{B}- 2\pi c_{IJ} \hat{\mathbf{I}}\cdot \hat{\mathbf{J}}\bigg)\text{,}
\end{equation}
where $\hat{\mathbf{J}}$ is the rotational angular momentum operator, $\hat{\mathbf{I}}$ is the nuclear spin operator for the proton,  $\mu_{N}$ is the nuclear magneton, $g_J$ is the g-factor for the rotational angular momentum for the $J$-th rotational manifold, $g_I$ is the proton g-factor, and $c_{IJ}$ is the spin-rotation constant~\cite{chou17}. To experimentally determine $g_J$ and $c_{IJ}$, we independently determine the magnetic field $|\mathbf{B}|$ by measuring the frequency difference between the $|S_{1/2}, m_j = -1/2\rangle\leftrightarrow|D_{5/2}, m_j = -5/2\rangle$ and $|S_{1/2}, m_j = -1/2\rangle\leftrightarrow|D_{5/2}, m_j = 3/2\rangle$ transitions of \Ca and backing out $|\mathbf{B}|$ from their well known dependence on this field ~\cite{roos99}. In addition, we measure the frequencies of three transitions between magnetic sublevels (at least one of them is the $|J,-J-1/2,-\rangle$ or $|J,-J+1/2,-\rangle$ states connected by the ``signature'' transition) within rotational manifolds with $J\in\{1,2,...,6\}$. The results are listed in Table~\ref{Table3}.  The coupling coefficients $g_J$ and $c_{IJ}$ (listed in Table~\ref{Table4}) can then be fitted from the measured transition frequencies and, along with the measured $|\mathbf{B}|$, be plugged in the Breit-Rabi formula for \CaH~\cite{chou17} to infer the frequency difference of the transition frequencies $|J,-J-1/2,-\rangle\leftrightarrow|J+2,-J+1/2,-\rangle$ measured in the laboratory and reported in Table~\ref{Table1} and between centroids that are relevant for defining the rotational constants in Eq.(\ref{E_J}). 
Subsequently, the frequency difference between the centroids can be derived and used to solve for the rotational constants $C_k$ in Eq.~(\ref{E_J}). With the frequencies of four transitions, we invert the linear system of equations for the rotational constants $C_k$ $\{k=1,2,3,4\}$ by setting $C_k = 0 $ for $k>4$. After determining the transition frequencies, we discovered a systematic level shift due to the radio-frequency (RF) electric field of the trap. We were unable to determine the precise amount of micro-motion during the measurements in hindsight, but we were able to conservatively constrain possible frequency shifts to the uncertainties given in the last column of Table I by additional measurements with deliberately exaggerated trap RF electric field.


\begin{table}[H]
\caption{The frequencies of in-manifold transitions for the $J = 1$ to $J=6$ rotational manifolds. The ``signature'' transition in each transition is listed in bold fonts. The 95~\% confidence intervals for the transitions are conservatively estimated at $\pm 1$~kHz.}
\begin{tabular}{ccc}
\\
\hline
\hline
\multicolumn{1}{c}{$J$} & \multicolumn{1}{c}{Transition} &\multicolumn{1}{c}{Frequency (kHz)}
\\
\hline
$1 $         &$\mathbf{|1,-3/2,-\rangle\leftrightarrow|1,-1/2,-\rangle}$& $\mathbf{9.9}$  
\\ $1 $         &$|1,-3/2,-\rangle\leftrightarrow|1,-1/2,+\rangle$& 9.1
\\ $1$         &$|1,-1/2,-\rangle\leftrightarrow|1,1/2,-\rangle$& 7.2
\\ $2 $         &$\mathbf{|2,-5/2,-\rangle\leftrightarrow|2,-3/2,-\rangle}$& $\mathbf{13.1}$
\\ $2$         &$|2,-5/2,-\rangle\leftrightarrow|2,-3/2,+\rangle$& 4.3
\\ $2 $        &$|2,-3/2,-\rangle\leftrightarrow|2,-1/2,-\rangle$& 7.4 
\\ $3 $        &$\mathbf{|3,-7/2,-\rangle\leftrightarrow|3,-5/2,-\rangle}$& $\mathbf{18.2}$
\\ $3 $        &$|3,-5/2,-\rangle\leftrightarrow|3,-3/2,-\rangle$& 7.0
\\ $3 $        &$|3,-5/2,-\rangle\leftrightarrow|3,-3/2,+\rangle$& 19.3
\\ $4 $        &$\mathbf{|4,-9/2,-\rangle\leftrightarrow|4,-7/2,-\rangle}$& $\mathbf{24.8}$
\\ $4$        &$|4,-7/2,-\rangle\leftrightarrow|4,-5/2,-\rangle$& 6.5
\\ $4$        &$|4,-7/2,-\rangle\leftrightarrow|4,-5/2,+\rangle$& 23.7
\\ $5 $        &$\mathbf{|5,-11/2,-\rangle\leftrightarrow|5,-9/2,-\rangle}$& $\mathbf{31.9}$
\\ $5$        &$|5,-9/2,-\rangle\leftrightarrow|5,-7/2,-\rangle$& 6.1
\\ $5$        &$|5,-9/2,-\rangle\leftrightarrow|5,-7/2,+\rangle$& 29.5
\\ $6$        &$\mathbf{|6,-13/2,-\rangle\leftrightarrow|6,-11/2,-\rangle}$& $\mathbf{39.4}$
\\ $6$        &$|6,-11/2,-\rangle\leftrightarrow|6,-9/2,-\rangle$& 5.6
\\ $6$        &$|6,-11/2,-\rangle\leftrightarrow|6,-9/2,+\rangle$& 35.9
\\
\hline
\hline
\end{tabular}
\label{Table3}
\end{table}

\begin{table}[H]
\caption{Inferred values of spin-rotation constant $c_{IJ}$ and g-factor $g_J$ for the $J = 1$ to $J=6$ rotational manifolds from the frequencies of three in-manifold transitions.}
\begin{tabular}{ccc}
\\
\hline
\hline
\multicolumn{1}{c}{$J$} & \multicolumn{1}{c}{$c_{IJ}$ (kHz)} &\multicolumn{1}{c}{$g_J$}
\\
\hline
   $1 $        & 8.2(1.4) & -1.39(0.44)  
\\ $2 $        & 8.06(65) & -1.39(36)  
\\ $3 $        & 8.04(25) & -1.36(24)  
\\ $4 $        & 8.08(19) & -1.37(23)  
\\ $5 $        & 8.08(15) & -1.38(22) 
\\ $6 $        & 8.06(13) & -1.37(22)  
\\
\hline
\hline
\end{tabular}
\label{Table4}
\end{table}

\subsection{Computational electronic structure methods}
The rotational constants of \CaH~were calculated
using a potential energy curve determined via
a composite scheme consisting of complete basis set extrapolated coupled-cluster calculations and incremental corrections for relativistic, and Born-Oppenheimer effects.
All calculations employed atom-centered Gaussian basis sets, 
the correlation-consistent polarized basis sets (cc-p)~\cite{RN272,RN273}, 
with valence-only (cc-pV) and core-valence (cc-pCV) correlation for Ca.
According to the usual nomenclature, the full basis set is specified by the 
number X of independent radial basis functions per correlated occupied orbital (XZ), for example cc-pCV5Z.
Due to an irregular basis-set convergence and extrapolation behavior observed,
the cc-pCVXZ (X = T, 4, 5) have been modified to improve systematic convergence to the basis set limit.
For Ca the s and p part of the latter basis sets have been replaced by the 
decontracted sp set of the cc-pCV5Z basis.
The resulting basis sets are denoted as cc-pCVXZ-mod (X = T, 4, 5).
Hartree-Fock (HF) and coupled-cluster singles and doubles 
augmented by a perturbative treatment of triple excitations (CCSD(T))~\cite{ccsdpt}
energies were computed using the decontracted cc-pCVXZ-mod (X = T, 4, 5)
basis sets for calcium and the decontracted aug-cc-pVXZ (X = T, 4, 5) basis sets for hydrogen.
This combination is indicated  by AUG-pCVXZ (X = T, 4, 5) in the following, see Table \ref{Table5} for details.
Complete basis set extrapolations were carried out separately for HF 
and correlation energies at the CCSD(T) level of theory, both with the AUG-pCVXZ basis sets.
For HF, an exponential formula from Ref. \cite{feller} is employed using always the three values obtained with X = T, 4, and 5.
The $l^{-3}$-formula \cite{helgaker} is employed for the correlation energy, using two values (either X = 3, 4 or X = 4, 5).
These two possibilities are denoted as CBS-34 and CBS-45.
To correct for electron correlation beyond CCSD(T),
incremental corrections at the 
coupled-cluster with singles, doubles, and triples (CCSDT) \cite{ccsdt} as well as 
the coupled-cluster with singles, doubles, triples and quadruples excitation (CCSDTQ) \cite{ccsdtq} level
were computed using cc-pV5Z and cc-pVTZ basis sets, respectively.
For valence-only basis sets, 
the frozen-core approximation has been used in coupled-cluster calculations, 
with the 5 lowest (doubly occupied) orbitals in the frozen core.
For basis sets including core-valence correlation, all electrons were correlated.
In the present work, 
Diagonal Born-Oppenheimer corrections (DBOC), 
which can be considered as a first-order correction to the
electronic energy associated with the nuclear kinetic energy operator,
are taken into account  at the CCSD/cc-pCVQZ level.~\cite{dboc}
Relativistic effects were calculated at the CCSD(T)/AUG-pCV5Z level
via a direct perturbation theory (DPT2).~\cite{dpt2}
All calculations were carried out using the quantum-chemical program package CFOUR~\cite{cfour} 
(Coupled-Cluster techniques for Computational Chemistry)
except those involving approximate coupled-cluster models including higher than
triple excitations, which were computed with the string-based many-body code
MRCC \cite{mrcc} interfaced to CFOUR.
In all cases, closed-shell restricted Hartree-Fock references were employed.

\begin{table}[H]
\caption{Overview of basis sets and contraction schemes used in this study.}
\begin{tabular}{c|ccc|ccc}
\hline 
\hline
Basis     & Basis set on Ca & primitives      & contraction & Basis set on H & primitives & contraction \\ 
\hline
AUG-pCVTZ & cc-pCVTZ-mod$^a$    & 26s18p8d2f      & no          & aug-cc-pVTZ & 6s3p2d     & no         \\  
AUG-pCVQZ & cc-pCVQZ-mod$^a$    & 26s18p10d4f2g   & no          & aug-cc-pVQZ & 7s4p3d2f   & no         \\  
AUG-pCV5Z & cc-pCV5Z-mod$^a$    & 26s18p12d6f4g2h & no          & aug-cc-pV5Z & 9s5p4d3f2g & no         \\ 
cc-pCVQZ  & cc-pCVQZ        & 25s19p10d4f2g   & 10s9p7d4f2g & cc-pVQZ     & 6s3p2d1f   & 4s3p2d1f   \\
cc-pV5Z   & cc-pV5Z         & 26s18p8d3f2g1h  & 8s7p5d3f2g1h& cc-pV5Z     & 8s4p3d2f1g & 5s4p3d2f1g \\
cc-pVTZ   & cc-pVTZ         & 20s14p6d1f      & 6s5p3d1f    & cc-pVTZ     & 5s2p1d     & 3s2p1d     \\
\hline
\hline
\end{tabular}
\label{Table5}
{$^a$cc-pCVXZ-mod (X = T, 4, 5) denotes a modified cc-pCVXZ (X = T, 4, 5) basis set for Ca, where the s and p part of the basis have been replaced by the decontracted sp set of the cc-pCV5Z basis set.}
\end{table}

\subsection{Computational determination of rotational constants}
Vibrational-rotational wave functions are obtained by numerical solution of the 1D Schr\"{o}dinger equation using the full potential $V_0(r)$ (energy of the electronic wave function as a function of Ca-H distance) plus centrifugal potential:
\begin{align}
V(r) = V_0(r) + \frac{\hbar^2 J(J+1)}{2\mu r^2},
\end{align}
where $\mu$ is the reduced mass.
The computed potential $V_0(r)$ is interpolated using cubic splines and the Schr\"{o}dinger equation is solved on an evenly spaced grid of 1500 points from $r$ = 1.0 $\mathrm{\AA}$ to 7.0 $\mathrm{\AA}$.
The rotational constants $C_k$ are obtained by fitting Eq.~\ref{E_J} of the main text to the computed energy levels up to $J=10$ and including up to the third centrifugal correction.

\begin{figure}

\includegraphics[angle=0, width=0.6\textwidth]{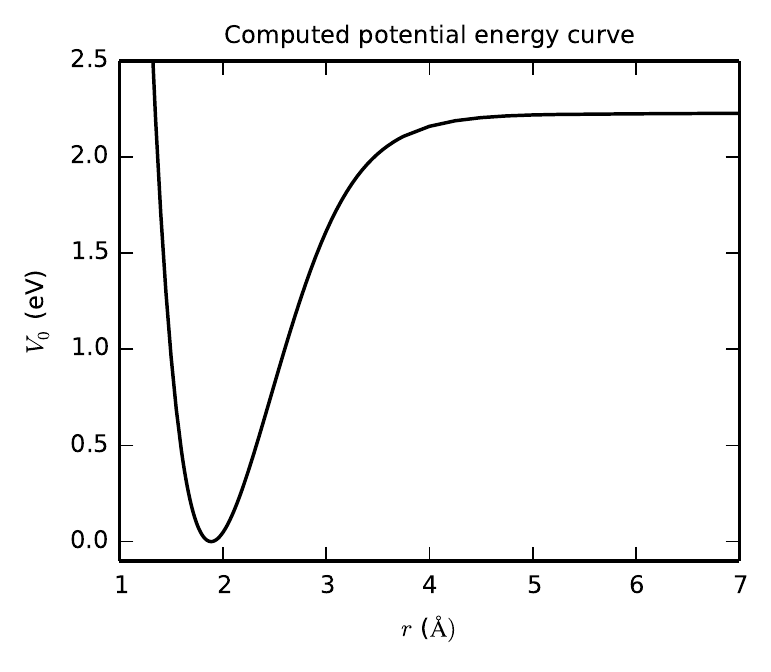}
\caption[Potential energy curve]{Computed potential energy curve at the CCSD(T)/CBS-45+$\Delta$DBOC+$\Delta$DPT2+$\Delta$T+$\Delta$Q level of theory relative to the energy minimum.
}
\label{PEC}
\end{figure}

For $B_R$, an additional correction is added according to~\cite{RN269}:
\begin{align}
\Delta B_\mathrm{el} = \frac{m_e}{m_p} \times g \times B_R\text{.}
\end{align}
With the rotational g tensor, $g \approx -1.35$, we find $\Delta B_\mathrm{el} = -0.105$ GHz.
Table \ref{tab_theory1} lists the value computed for $B_R$ at increasing levels of accuracy.
The largest effect on $B_R$ is due to basis set size with variations on the order of 1 GHz.
All other effects, such as an improved description of electron correlation by going from CCSD(T) to CCSDT and CCSDTQ as well as the description of relativistic effects and the diagonal Born-Oppenheimer correction are only on the order of 0.1 GHz.
For the incremental corrections, basis set convergence was also checked.
Generally, deviations of the derived corrections varied by less than 0.02 GHz if at least a cc-pVTZ basis set is used.
The only exception is the relativistic correction, where larger deviations are observed.
Here, the deviation between the MVD1-correction evaluated using the AUG-cc-pCVXZ basis sets with X = 4, 5 is 0.02 GHz, 
while the deviation to the cc-pVXZ basis sets with X = T, 4, 5 is up to 0.1 GHz.
When computing relativistic corrections using mass-velocity-1-electron-Darwin (MVD1) instead of DPT2 corrections, effects on the rotational constants are small ($\approx$ 0.01 GHz). The best potential energy curve is shown in Fig.~\ref{PEC}. Out of computed values for $B_R$ in the literature~\cite{abe10,abe12,condoluci17}, $B_R$ = 143.8 GHz is at the most accurate level of theory (CCSDT/cc-pCV5Z)~\cite{condoluci17}.

\begin{table}[H]
\caption{$C_1$ (including the correction $\Delta B_\mathrm{el} = -0.11$ GHz) and the higher rotational constants at various levels of theory with increasing accuracy.
Computation with $\Delta$DPT2:CCSD(T)/AUG-pCV5Z, $\Delta$DBOC:CCSD/cc-pCVQZ, $\Delta$T/cc-pV5Z and $\Delta$Q/cc-pVTZ.
Complete basis set extrapolation (CBS-34, CBS-45) is done with the AUG-pCVXZ basis sets as explained in the main text.}
\begin{tabular}{lrrrr}
\\
\hline
\hline
Method & $C_1$ (GHz) &  $C_2$ (MHz) &  $C_3$ (Hz) &  $C_4$ (Hz)
\\
\hline
 CCSD(T)/AUG-pCVTZ                                            &140.56&	-5.801&	223.4&	-0.0155\\
 CCSD(T)/AUG-pCVQZ                                            &141.97&	-5.829&	227.7&	-0.0158\\
 CCSD(T)/AUG-pCV5Z                                            &142.50&	-5.839&	227.4&	-0.0159\\
 CCSD(T)/CBS-34                                               &142.90&	-5.848&	231.7&	-0.0161\\
 CCSD(T)/CBS-45                                               &143.02&	-5.848&	227.1&	-0.0161\\
 CCSD(T)/CBS-45+$\Delta$DBOC                                  &142.92&	-5.839&	226.3&	-0.0161\\
 CCSD(T)/CBS-45+$\Delta$DBOC+$\Delta$DPT2                     &142.73&	-5.829&	222.6&	-0.0158\\
 CCSD(T)/CBS-45+$\Delta$DBOC+$\Delta$DPT2+$\Delta$T           &142.67&	-5.831&	222.4&	-0.0158\\
 CCSD(T)/CBS-45+$\Delta$DBOC+$\Delta$DPT2+$\Delta$T+$\Delta$Q &142.69&	-5.831&	222.6&	-0.0158\\
\hline
\hline
\end{tabular}
\label{tab_theory1}
\end{table}

To estimate the uncertainties of the different components 
of the theoretical composite scheme applied here, 
absolute differences 
of consecutive members of systematic series of methods
and basis sets are used as estimates of one half of 95~\% confidence limits.
As already mentioned, the HF+CCSD(T) contribution shows the largest variation,
taking the difference between the plain CCSD(T)/AUG-pCV5Z and CCSD(T)/CBS-45 would result in 
estimates for 95~\% confidence limits of $\pm$ 0.42 GHz, 0.01 MHz, 0.3 Hz, and less than 0.001 Hz for 
$C_1$, $C_2$, $C_3$, and $C_4$, respectively.
Considering the facts that in the basis sets employed 
the s and p basis functions are kept constant on calcium
and that the effect of basis set extrapolation is larger than the difference between both
results with the basis sets used for the extrapolation for $C_1$, 
the estimates for the 95~\% confidence limits
have been augmented to the absolute difference between the results 
of CCSD(T)/AUG-pCVQZ and CCSD(T)/CBS-45.
The 95~\% confidence limits of the additional contributions 
$\Delta$DBOC, $\Delta$DPT2, $\Delta$T, and $\Delta$Q are obtained as differences 
to results with smaller basis sets and are found to be almost insignificant.

\begin{table}[H]
\caption{Uncertainty estimates for the different contributions.
Computation with $\Delta$DPT2:CCSD(T)/AUG-pCV5Z, $\Delta$DBOC:CCSD/cc-pCVQZ, $\Delta$T/cc-pV5Z and $\Delta$Q/cc-pVTZ.
Complete basis set extrapolation (CBS-34, CBS-45) is based on the AUG-pCVXZ basis sets as explained in the main text.}
\begin{tabular}{lrrrl}
\\
\hline
\hline
Method & $\delta C_1$ (GHz) &  $\delta C_2$ (MHz) &  $\delta C_3$ (Hz) &  $\delta C_4$ (Hz)
\\
\hline
CCSD(T)/CBS-45                                               & $\pm$1.05 & $\pm$0.019& $\pm$0.6 & $\pm$0.0004\\
$\Delta$DBOC                                                 & $\pm$0.00 & $\pm$0.000& $\pm$0.0 & $\pm$0.0000$^a$\\
$\Delta$DPT2                                                 & $\pm$0.01 & $\pm$0.001& $\pm$0.0 & $\pm$0.0000$^a$\\
$\Delta$T                                                    & $\pm$0.01 & $\pm$0.002& $\pm$0.0 & $\pm$0.0000$^a$\\
$\Delta$Q                                                    & $\pm$0.02 & $\pm$0.000& $\pm$0.2 & $\pm$0.0000$^a$\\
\hline
CCSD(T)/CBS-45+$\Delta$DBOC+$\Delta$DPT2+$\Delta$T+$\Delta$Q$^b$& $\pm$1.05 & $\pm$0.019 &$\pm$0.6&$ \pm$0.0004\\
\hline
\hline
\multicolumn{5}{l}{$^a$  0.0000 denotes an uncertainty smaller than 0.00005.}\\
\multicolumn{5}{l}{$^b$  Combined uncertainties were obtained, using summation in quadrature.}\\
\end{tabular}
\label{tab_unc}
\end{table}

\end{document}